\documentclass[a4paper]{article}
\usepackage[n, advantage, operators,sets, adversary,landau,probability,notions,logic,ff,mm,primitives,events, complexity,asymptotics,keys]{cryptocode}
\usepackage{geometry}
\usepackage{enumerate}
\usepackage{amsmath}
\usepackage{mathtools}
\usepackage[english]{babel}
\usepackage{hyperref}
\usepackage{tikz}
\usepackage{url}
\usepackage{datetime2}
\usepackage{float}
\hypersetup{
	linktoc=all
}

\usepackage{listings}
\usepackage{color}
\definecolor{blue}{rgb}{0.04, 0.06, 0.66}
\definecolor{green}{rgb}{0.12, 0.58, 0}
\usepackage{listings}
\usepackage{fancyhdr}
\title{\Huge 00}
\author{Nguyen Thoi Minh Quan 
\footnote{https://www.linkedin.com/in/quan-nguyen-a3209817,
https://scholar.google.com/citations?user=9uUqJ9IAAAAJ, https://github.com/cryptosubtlety, msuntmquan@gmail.com}
\footnote{Disclaimer: This is my personal research, and hence it does not represent the views of my employer.}}

\begin{document}
\date{}
\maketitle
\begin{abstract}
\normalsize
What is the funniest number in cryptography (\emph{Episode 2})? 0 \cite{0}. The reason is that $\forall x, x \cdot 0 = 0$, i.e., the equation is satisfied no matter what $x$ is. We'll use zero to attack zero-knowledge proof (ZKP). In particular, we'll discuss a \emph{critical} issue in a cutting-edge ZKP PLONK \cite{plonk} C++ implementation which allows an attacker to create a \emph{forged} proof that all verifiers will accept. We’ll show how theory guides the attack’s direction. In practice, the attack works like a charm and we'll show how the attack falls through \emph{a chain of perfectly aligned software cracks.}

In the same codebase, there is an \emph{independent critical} ECDSA bug where (r, s) = (0, 0) is a valid signature for arbitrary keys and messages, but we won’t discuss it further because it’s a known ECDSA attack vector in the Google Wycheproof cryptanalysis project \cite{wycheproof} that I worked on a few years ago.

All bugs have been responsibly disclosed through the vendor's bug bounty program with total reward $\sim$ \$15,000 (thank you).

\end{abstract}

\section*{How theory guides the attack's direction?}
In any zero-knowledge proof (ZKP) \footnote{There are subtle differences between proof vs argument; soundness vs knowledge soundness; and zero knowledge but we won't need those details in this article.} system, there is a prover and a verifier. The prover has to convince the verifier that it knows witness of a certain statement. The verifier has to check whether a certain equation is satisfied or not. PLONK uses polynomial commitment\cite{kzg10}, and pairing \cite{benlynnnote}\cite{cryptobook}. For the purpose of this article, you don’t have to know what polynomial commitment or pairing $e$ is. All you need to know is that the pairing $e(P_1, P_2)$ maps 2 points $P_1, P_2$ to a finite field and $G_1, G_2$ are 2 base points on the elliptic curves.

One distinctive notation in PLONK is the following: $[x]_1 = x \cdot G_1, [x]_2 = x \cdot G_2$. What it means is that when $[x]_1$ is published, the attacker does not know $x$, the attacker only knows the product $x\cdot G_1$ and without breaking discrete log problems, $x$ remains secret. \emph{However, the attacker can manipulate $[x]_1$}. We’ll use this observation in the attack.

The final step in the verifier is to verify following equation
\begin{align*}
e([W_z]_1 + u\cdot [W_{z\omega}]_1, [x]_2)\cdot e(-(z\cdot [W_z]_1 + uz\omega\cdot [W_{z\omega}]_1 + [F]_1 - [E]_1), [1]_2) \stackrel{?}= 1
\end{align*}
where 
\begin{figure}[H]
\includegraphics[width=\linewidth]{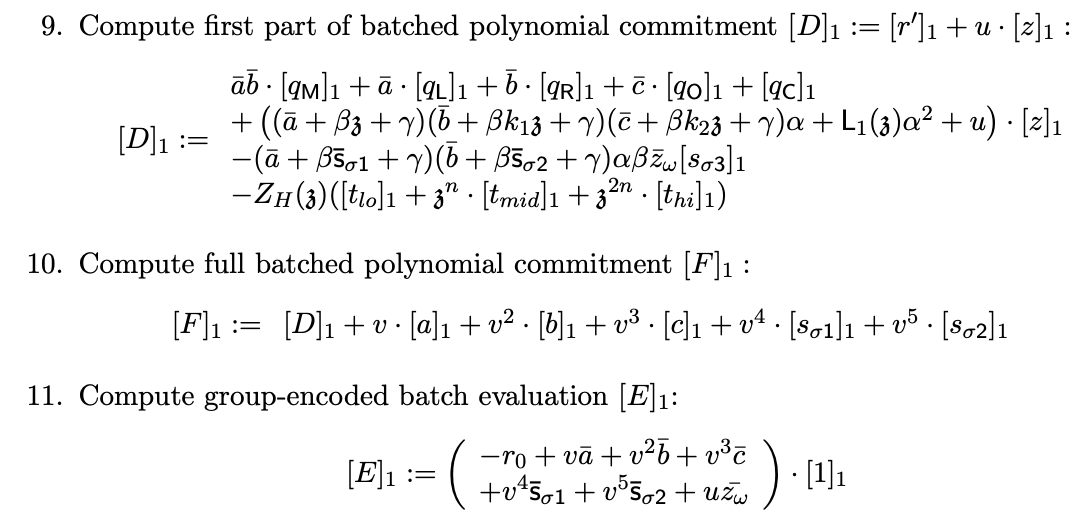}
\end{figure}

The formulas look scary, don't they? Just count how many parameters there are. To simplify it,  we’ll denote 
\begin{align*}
P[1] &= [W_z]_1 + u\cdot [W_{z\omega}]_1 \\
P[0] &= -(z\cdot [W_z]_1 + uz\omega \cdot [W_{z\omega}]_1 + [F]_1 - [E]_1)
\end{align*}
and the equation becomes
\begin{align*}
e(P[1], [x]_2) \cdot e(P[0], [1]_2) \stackrel{?}= 1
\end{align*}
Now, it’s simple. I'm kidding.

When dealing with such complexity, it’s important to find the weakest and easiest place to
attack. Recall that we want a full verification bypass, i.e., the attacker doesn’t know any witness, but wants to convince the verifier to accept the proof. Therefore, I was looking for 2 things:
\begin{enumerate}
\item Parameters that attackers can manipulate.
\item The least effort way to manipulate parameters. 
\end{enumerate}
while forcing the verification equation satisfied.

To give you an brief idea:
\begin{enumerate}[$\diamondsuit$]
\item $[W_z]_1, [W_{z\omega}]_1$ are under the attacker's control. To be clear, the attacker can manipulate $[W_z]_1, [W_{z\omega}]_1$ but the attacker does not know the inside true values $W_z, W_{z\omega}$.
\item $u$ = hash(transcript) where hash acts as a random oracle, so it’s technically outside the attacker's control.
\item $x$ is part of a trusted setup that no one (including the prover and verifier) is assumed to know.
\item $F$ and $E$ are computed by the verifier (not attacker) in a complicated multi-steps process, so let’s ignore them.
\end{enumerate}

All right, it seems that $[W_z]_1, [W_{z\omega}]_1$ are obvious targets to manipulate. What if the attacker uses $[W_z]_1 = 0, [W_{z\omega}]_1 = 0$ where $0$ is an identity (infinity) point in the elliptic curve?
\begin{enumerate}[$\diamondsuit$]
\item $P[1] = [W_z]_1 + u\cdot [W_{z\omega}]_1 = 0 + u\cdot 0 = 0$ independent of the transcript's hash $u$ (aka Fiat-Shamir transform \cite{FiatShamir}).

\emph{Basically, we neutralize the role of Fiat-Shamir transform}.

Now, there is some hope!
\item $e(P[1], [x]_2)  = e(0, [x]_2) = 1$ because $e(0, R) = e(R, 0) = 1, \forall R$.
\item We have \begin{align*}
e(P[0], [1]_2) &= e(-(z\cdot[W_z]_1 + uz\omega \cdot [W_{z\omega}]_1 + [F]_1 - [E]_1), [1]_2)  \\
&= e(-(z\cdot 0 + uz\omega \cdot 0 + [F]_1 - [E]_1), [1]_2)  \\
&= e(-(0 + 0 + [F]_1 - [E]_1), [1]_2) \\
&= e(-([F]_1 - [E]_1), [1]_2) \\
& \neq 1
\end{align*}
because $[F]_1 - [E]_1 \neq 0$.

\end{enumerate}
Therefore, $e(P[1], [x]_2)\cdot e(P[0], [1]_2) = 1 \cdot e(P[0], [1]_2) \neq 1$. So, the attack doesn't work?

\section*{Why does the attack work in practice?}
Fortunately, I’m not a theoretical cryptographer. I don’t believe in theory’s security, at least not
completely. I trust the program, the binary that runs. Whatever the program output tells me,
that’s the truth. So I just input $[W_z]_1 = 0, [W_{z\omega}]_1 = 0$ to the PLONK’s verifier and see what would happen. The verifier computes $e(P[1], [x]_2)\cdot e(P[0], [1]_2) = 1$ and returns true, so the attack completely bypasses the verifier. Woo-hoo!

Note that it’s not strange at all in the attack process where the attacker observes unexplainable
behavior. It’s pretty normal and common. Sometimes, it’s the start of a surprise attack. The investigation showed that
the attack falls through \emph{a chain of perfectly aligned software cracks}.

\subsection*{The root cause}
Before we move on, a technical
software implementation detail is that points in the elliptic curve are often represented in 3
forms: byte array (on the wire or storage), affine coordinate $P = (x, y)$ or projective coordinate $P = (x, y, z)$.

Recall that in our attack vector, we use $[W_z]_1 = 0 = (0, 0)$, $[W_{z\omega}]_1 = 0 = (0, 0)$ or $(P[0] \neq 0, P[1] = 0)$ where $0 = (0, 0)$ means its affine coordinate $(x, y) = (0, 0)$. As a reminder, the attackers don't control $P[0], P[1]$, they have to manipulate them through $[W_z]_1
, [W_{z\omega}]_1$. If you’re lazy, just
use a zero byte array for the whole proof (which includes $[W_z]_1
,[W_{z\omega}]_1$) and it will bypass all
verifiers.

\begin{enumerate}
\item  The verifier checks whether $[W_z]_1, [W_{z\omega}]_1$ are on the elliptic curve or not. $[W_z]_1, [W_{z\omega}]_1$ are \emph{not} valid points on
the curve. However, the verifier \emph{does not stop} immediately when it sees invalid points. It
continues the execution, but it excludes the invalid 0 points in \emph{some} further computations. The
amazing thing is that while $[W_z]_1 = 0, [W_{z\omega}]_1 = 0, P[1] = 0$ are excluded in some further computations, they're included in the crucial computation with pairing which allows the attack to work. If the program returns
false immediately once it sees invalid points, the attack would fail.

\item In the elliptic curve code, there is another check to \emph{reject the infinity point}. However,
according to the code, $P[1] = 0$ is not infinity. The infinity method checks whether
the most significant bit of the $P[1]$ is 1, but P[1] = 0’s most significant bit is 0. Hence $P[1] = 0$
bypasses the infinity point check.

\item In the computation process, there is a method that computes the inverse of 0 mod p. The method doesn't check for 0 input. It uses Fermat's little theorem, i.e., it uses equation $x^{p - 1} = 1 \mod p$ or $x^{p - 2} \cdot x = 1 \mod p$ or $x^{p - 2}$ is the inverse of $x \mod p$. However, when $x = 0$, $x^{p - 2} = 0$ which means that the inverse of $0 \mod p$ is 0. This isn't even correct mathematically because the inverse of 0 mod p shouldn’t exist. If the inverse method rejects 0, the attack would again fail.

\item  Now, the array $(P[0], P[1]) = (P[0] \neq 0, P[1] = 0)$ are in projective coordinates $(x, y, z)$. The
projective coordinate is often just an intermediate representation for optimization purposes. After
finishing the computation, the projective coordinates go through a process called normalization to
eliminate $z$ (i.e. to make $z = 1$) and goes back to affine coordinate $(x, y) \sim (x, y, 1)$. The code does not normalize points
individually, instead it \emph{batch-normalizes} an array of points together where $P[1].z = 0$ will affect
$P[0]$. The vulnerable code outputs $(P[0], P[1]) = (0, 0)$, i.e., \emph{it turns non-zero point P[0] into a 0
point}. For instance, here is the output from the verifier
\begin{verbatim}
"Before batch_normalize
P[0]: { 0x12270675066dbf202e8766f5fa48648f95032fbff46996a08e05e427ed0fffb9,
0x2cce89ca786bd0a3db55776a24aa3253bce3b8ef689849f93596b5b26afec90f,
0x04ae1f4cd5f84a484acc4ba115fbd02a879d2e30b8cd97e18f3865887213823b }
P[1]: { 0x0000000000000000000000000000000000000000000000000000000000000000,
0x0000000000000000000000000000000000000000000000000000000000000000,
0x0000000000000000000000000000000000000000000000000000000000000000 }
After batch_normalize
P[0]: { 0x0000000000000000000000000000000000000000000000000000000000000000,
0x0000000000000000000000000000000000000000000000000000000000000000,
0x0000000000000000000000000000000000000000000000000000000000000001 }
P[1]: { 0x0000000000000000000000000000000000000000000000000000000000000000,
0x0000000000000000000000000000000000000000000000000000000000000000,
0x0000000000000000000000000000000000000000000000000000000000000001 }"
\end{verbatim}
It's worth noting in projective coordinates (x, y, z), we should reject z = 0 for regular
points, but the code doesn't. If the code rejects z = 0 in projective coordinates, the attack would
again fail. As a small note, it’s worth noting that if each point is normalized independently then
after normalization P[1] would be different from zero causing the attack to fail.
As a consequence the final verifier’s computation becomes $e(P[1], [x]_2) \cdot e(P[0], [1]_2) = e(0, [x]_2)
\cdot e(0, [1]_2)$.

\item Lastly, while $P[1] = 0$ is not on the curve and $P[1] = 0$ is \emph{not} infinity according to step 2, the
\emph{pairing} code considers P[1] = 0 as infinity, in the sense that $e(0, R) = 1, \forall R$. Without these
\emph{contradicting} views of the same input 0 in step 2 and step 5, the attack would fail.
\end{enumerate}

\section*{Acknowledgements}
We thank Ariel Gabizon for the feedback on this article.
\newpage
\bibliographystyle{unsrt}
\bibliography{research}
\end{document}